\documentclass[aps,prl,preprintnumbers,superscriptaddress]{revtex4-2}

\usepackage{epsfig}
\usepackage{amssymb}
\usepackage{times}
\usepackage{amsmath}
\usepackage{color}

\begin{document}

\preprint{\large DESY-26-125}

\bigskip

\title{Generation of 10-TW Attosecond X-Ray Pulses in a Free Electron Laser}



\author{E.A.~Schneidmiller}\email{evgeny.schneidmiller@desy.de} \affiliation{\small Deutsches Elektronen-Synchrotron DESY, Notkestr. 85, 22607 Hamburg, Germany}
\author{S.~Serkez} \affiliation{\small European XFEL, Holzkoppel 4, 22869 Schenefeld, Germany}
\author{ I.~Zagorodnov} \affiliation{\small Deutsches Elektronen-Synchrotron DESY, Notkestr. 85, 22607 Hamburg, Germany}
\author{N.~Gerasimova} \affiliation{\small European XFEL, Holzkoppel 4, 22869 Schenefeld, Germany}
\author{M.~Scholz}  \affiliation{\small Deutsches Elektronen-Synchrotron DESY, Notkestr. 85, 22607 Hamburg, Germany}
\author{N.~Golubeva}  \affiliation{\small Deutsches Elektronen-Synchrotron DESY, Notkestr. 85, 22607 Hamburg, Germany}
\author{M.~Guetg}  \affiliation{\small Deutsches Elektronen-Synchrotron DESY, Notkestr. 85, 22607 Hamburg, Germany}
\author{G.~Geloni} \affiliation{\small European XFEL, Holzkoppel 4, 22869 Schenefeld, Germany}
\author{S.~Karabekyan} \affiliation{\small European XFEL, Holzkoppel 4, 22869 Schenefeld, Germany}
\author{F.~Sottocorona}  \affiliation{\small Deutsches Elektronen-Synchrotron DESY, Notkestr. 85, 22607 Hamburg, Germany}
\author{A.~Trebushinin} \affiliation{\small European XFEL, Holzkoppel 4, 22869 Schenefeld, Germany}
\author{G.~Perosa} \affiliation{\small European XFEL, Holzkoppel 4, 22869 Schenefeld, Germany}
\author{S.~Tomin}  \affiliation{\small Deutsches Elektronen-Synchrotron DESY, Notkestr. 85, 22607 Hamburg, Germany}
\author{F.~Mayet}  \affiliation{\small Deutsches Elektronen-Synchrotron DESY, Notkestr. 85, 22607 Hamburg, Germany}
\author{Th.~Maltezopoulos} \affiliation{\small European XFEL, Holzkoppel 4, 22869 Schenefeld, Germany}
\author{J.~Gr{\"u}nert} \affiliation{\small European XFEL, Holzkoppel 4, 22869 Schenefeld, Germany}
\author{J.~Laksman} \affiliation{\small European XFEL, Holzkoppel 4, 22869 Schenefeld, Germany}
\author{H.~Ahmadi Rashtabadi} \affiliation{\small European XFEL, Holzkoppel 4, 22869 Schenefeld, Germany}
\author{T.M.~Baumann} \affiliation{\small European XFEL, Holzkoppel 4, 22869 Schenefeld, Germany}
\author{R.~Boll} \affiliation{\small European XFEL, Holzkoppel 4, 22869 Schenefeld, Germany}
\author{F.~Calegari}  \affiliation{\small Deutsches Elektronen-Synchrotron DESY, Notkestr. 85, 22607 Hamburg, Germany}
\author{G.~Catacchio}  \affiliation{\small Deutsches Elektronen-Synchrotron DESY, Notkestr. 85, 22607 Hamburg, Germany}
\author{S.~Dold} \affiliation{\small European XFEL, Holzkoppel 4, 22869 Schenefeld, Germany}
\author{D.E.~Ferreira de Lima} \affiliation{\small European XFEL, Holzkoppel 4, 22869 Schenefeld, Germany}
\author{L.~Funke}  \affiliation{\small Department of Physics, Center for Synchrotron Radiation (DELTA), Maria-Goeppert-Mayer-Str. 2, 44227 Dortmund, TU Dortmund}
\author{W.~Helml}  \affiliation{\small Department of Physics, Center for Synchrotron Radiation (DELTA), Maria-Goeppert-Mayer-Str. 2, 44227 Dortmund, TU Dortmund}
\author{M.~Ilchen}  \affiliation{\small Deutsches Elektronen-Synchrotron DESY, Notkestr. 85, 22607 Hamburg, Germany}
\affiliation{\small Institut f{\"u}r Experimentalphysik, Universit{\"a}t Hamburg, Luruper
Chaussee 149, 22761 Hamburg, Germany}
\author{T.~Mazza} \affiliation{\small European XFEL, Holzkoppel 4, 22869 Schenefeld, Germany}
\author{M.~Meyer} \affiliation{\small European XFEL, Holzkoppel 4, 22869 Schenefeld, Germany}
\author{T.~Mullins}  \affiliation{\small Deutsches Elektronen-Synchrotron DESY, Notkestr. 85, 22607 Hamburg, Germany}
\author{M.~Robinson} \affiliation{\small European XFEL, Holzkoppel 4, 22869 Schenefeld, Germany}
\author{S.~Savio}  
\affiliation{\small Institut f{\"u}r Experimentalphysik, Universit{\"a}t Hamburg, Luruper
Chaussee 149, 22761 Hamburg, Germany}
\affiliation{\small Department of Physics, Center for Synchrotron Radiation (DELTA), Maria-Goeppert-Mayer-Str. 2, 44227 Dortmund, TU Dortmund}
\author{K.~Scharei} \affiliation{\small Intelligent Embedded Systems, University of Kassel, Wilhelmsh{\"o}her Allee 71-73, 34121 Kassel, Germany}
\author{Ph.~Schmidt} \affiliation{\small European XFEL, Holzkoppel 4, 22869 Schenefeld, Germany}
\author{A.~Thiel}  \affiliation{\small Deutsches Elektronen-Synchrotron DESY, Notkestr. 85, 22607 Hamburg, Germany}
\affiliation{\small Institut f{\"u}r Experimentalphysik, Universit{\"a}t Hamburg, Luruper
Chaussee 149, 22761 Hamburg, Germany}
\author{S.~Usenko} \affiliation{\small European XFEL, Holzkoppel 4, 22869 Schenefeld, Germany}
\author{L.~W{\"u}lfing}  \affiliation{\small Department of Physics, Center for Synchrotron Radiation (DELTA), Maria-Goeppert-Mayer-Str. 2, 44227 Dortmund, TU Dortmund}
\author{W.~Decking} \affiliation{\small Deutsches Elektronen-Synchrotron DESY, Notkestr. 85, 22607 Hamburg, Germany}


\date{\today}

\begin{abstract}
Attosecond pulses provide direct access to electron dynamics in atoms, molecules, and solids on their natural timescales. While table-top sources have achieved attosecond pulse durations at moderate intensities, extending attosecond science to high peak powers in the x-ray regime requires free-electron lasers (FELs). Here we report the generation of soft-x-ray pulses with a duration of $\sim\!100$ as and peak power on the order of 10 TW at the European XFEL. The unprecedented combination of pulse duration and peak power is enabled by operation of a soft-x-ray undulator at high electron energy together with five-stage bunch compression, producing ultrahigh peak currents. These millijoule-level attosecond x-ray pulses establish a new regime for attosecond science.
\end{abstract}


\maketitle

The ability to generate attosecond pulses has opened direct access to electron dynamics in atoms, molecules, and solids on their natural timescales. This frontier of ultrafast science was recognized with the 2023 Nobel Prize in Physics \cite{nobel}, highlighting the foundational role of attosecond light in probing electronic motion. High-harmonic generation (HHG) sources have enabled remarkable advances in attosecond science \cite{krausz,calegari,biegert}. However, table-top HHG sources are intrinsically limited in pulse energy and peak power, restricting access to nonlinear and strongly driven regimes, particularly at short wavelengths.

The advent of free-electron lasers  \cite{book} has dramatically changed this landscape, pushing attosecond pulses toward shorter wavelengths in the x-ray range and higher pulse energies (up to hundreds of microjoules), thereby opening new research opportunities in this regime \cite{nora}. Following early theoretical proposals \cite{attoharm,atto100GW,fawley,esase,chirp-taper}, attosecond x-ray pulses are routinely generated at several x-ray FEL facilities \cite{lcls-hard,duris-lcls-soft,swiss,harmonics,franz-lcls-soft,yan,funke}. These facilities utilize the self-amplified spontaneous emission (SASE) mechanism \cite{part-acc}, in which initial shot-noise density fluctuations in the relativistic electron beam are exponentially amplified through interaction with the emitted electromagnetic field in long periodic magnetic structures (undulators). At the end of the undulator, saturation is reached, and the strongly microbunched electron beam radiates an intense, quasi-monochromatic pulse with wavelength

\begin{equation}
\lambda = \frac{\lambda_{\mathrm{w}} (1+K^2/2)}{2 \gamma^2}    
\label{lambda}
\end{equation}

\noindent where $\lambda_{\mathrm{w}}$ is the undulator period, $\gamma$ is the relativistic factor, and $K$ is the undulator parameter, proportional to the product of the undulator period and the magnetic field amplitude. For successful lasing, the electron bunch must have sufficiently high quality \cite{book}, i.e., a combination of high peak current $I$, relatively small uncorrelated energy spread $\sigma_{\mathcal{E}}/\mathcal{E}_0$, and low normalized emittance $\epsilon_n = \gamma \epsilon$. The latter parameter characterizes the transverse phase space of the electron beam, and the geometrical emittance $\epsilon$ typically satisfies the condition $\epsilon < \lambda/4\pi$.

For the generation of attosecond pulses, an additional requirement must be satisfied: the electron bunch, or at least its lasing fraction, must be sufficiently short. A natural timescale for comparison is the FEL coherence time, determined by the slippage of the radiation with respect to the electron bunch over one FEL gain length, i.e., the $e$-folding length of the amplified electromagnetic field \cite{book}. In this regime, light pulses containing a single dominant spike in both the temporal and spectral domains are generated with high probability, interleaved with stochastic shots exhibiting a few spikes (or modes).

\begin{figure*}[t]
\centering
\includegraphics[width=0.9\textwidth]{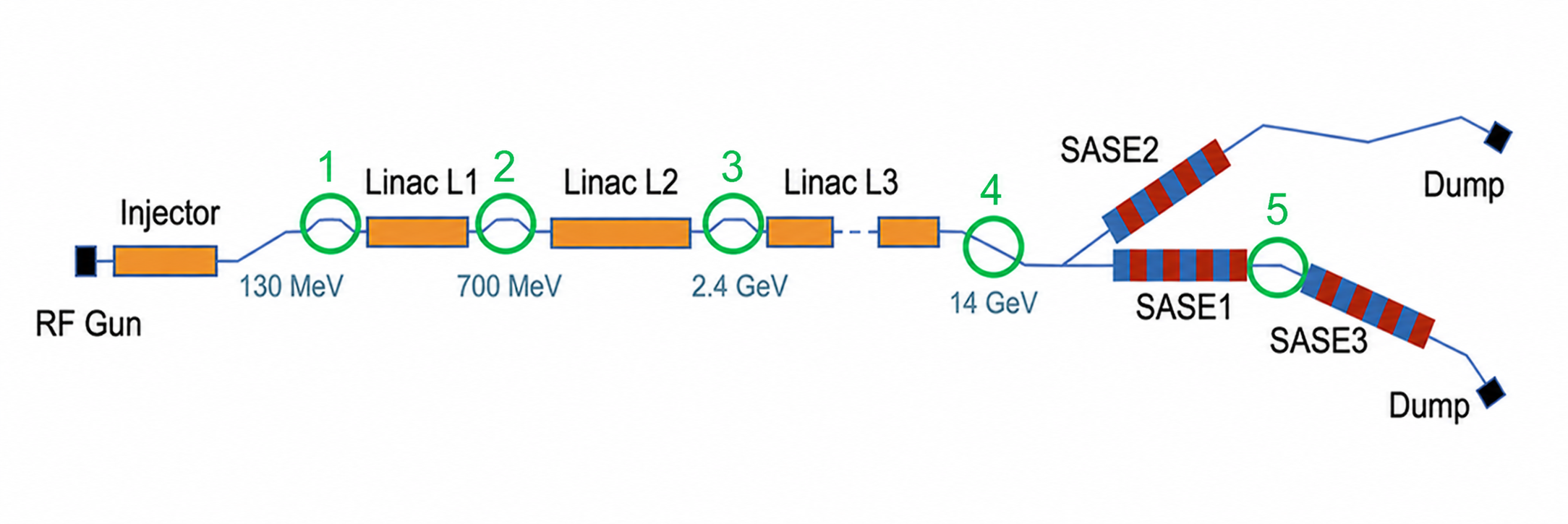}
\caption{
Layout of the European XFEL. The five compression stages employed in this work are indicated by green circles.
}
\label{XFEL_layout}
\end{figure*}

In this Letter, we report the generation of attosecond x-ray pulses with unprecedented properties at the European XFEL, a high-repetition-rate x-ray FEL user facility driven by a superconducting accelerator \cite{winni}. The unique layout of the facility and the distinctive parameters of its soft-x-ray undulator line allow extremely strong bunch compression while satisfying the conditions for lasing discussed above. The accelerator layout is shown in Fig.~\ref{XFEL_layout}. Low-emittance electron bunches with low initial peak current are generated in a laser-driven radio-frequency (RF) gun, accelerated to 130 MeV, and injected into the main linac. There, the beam is accelerated to its final energy and compressed in three magnetic chicanes (labeled 1–3 in Fig.~\ref{XFEL_layout}). During user operation, the accelerator provides several standard beam energies up to 16.3 GeV. In our experiments, the energy was 14 GeV. After passing a dogleg-collimator \cite{balandin} (labeled 4 in Fig.~\ref{XFEL_layout}), the beam is distributed between the two hard x-ray undulators, SASE1 and SASE2. The soft x-ray undulator SASE3 is located downstream of SASE1 and is decoupled from its operation by applying orbit kicks to selected bunches upstream of SASE1 with fast kickers and restoring the orbit downstream using DC steerers \cite{brinkmann}.

The key element of our method for generating intense attosecond pulses is the extremely strong compression of electron bunches. We illustrate this process using beam dynamics simulations for the machine parameters employed in the experiment. The longitudinal transformation through a dispersive section can be described by \cite{brown}

\begin{equation}
z_{f} = z_{i} + R_{56} \delta + T_{566} \delta^2 + O (\delta^3) ,
\label{comp_trans}
\end{equation}

\noindent where $z_{i}$ and $z_{f}$ denote the initial and final longitudinal positions relative to a reference particle, and $\delta$ is the relative momentum deviation. For ultrarelativistic beams, $\delta$ can equivalently be expressed as the relative energy deviation, $\delta = (\mathcal{E}-\mathcal{E}_0)/\mathcal{E}_0$. The coefficients $R_{56}$ and $T_{566}$ are the first- and second-order momentum compaction factors (longitudinal dispersions), with $R_{56}$ values for the three chicanes being positive \footnote{The sign convention follows Ref.~\cite{brown}.}.
In a typical compression scheme, a $z$–$\delta$ correlation (energy chirp) is introduced by off-crest acceleration, and the compression is linearized using a third-harmonic RF module in the injector. It is furthermore required that the head of the bunch has lower energy than the tail. The beam is typically undercompressed; i.e., in the final bunch compressor it does not reach an “upright” orientation in longitudinal phase space ($z$–$\delta$), and the typical peak current during user operation is about 5 kA at a beam energy of 2.4 GeV in BC2. 

To generate extremely intense attosecond x-ray pulses in SASE3, we substantially modify the compression scheme. First, instead of undercompression, we employ full compression of a relatively high-charge bunch (350 pC) in the last bunch compressor BC2 (labeled 3 in Fig.~\ref{XFEL_layout}). 

Second, we utilize the collimator as a fourth compression stage (labeled 4 in Fig.~\ref{XFEL_layout}). The fully compressed bunch exiting BC2 is subjected to strong collective effects downstream, most notably longitudinal space charge (LSC) in the linac section following BC2 and coherent synchrotron radiation (CSR) \cite{we-csr} in the collimator dipoles. These effects induce a positive $z$--$\delta$ correlation, corresponding to an energy chirp opposite in sign to the RF-induced chirp. In order to compress the bunch, we therefore require a momentum compaction of opposite sign. However, the linear momentum compaction of the collimator is essentially zero under nominal operation, precluding first-order compression. At the same time, the collimator possesses a significant second-order longitudinal dispersion, $T_{566}$, which can be exploited by introducing a finite energy offset $\delta_0$ of the reference particle with respect to the design energy. Expanding Eq.~(\ref{comp_trans}) about $\delta_0$ yields an effective linear momentum compaction,
$R_{56}^{\mathrm{eff}} \simeq 2\delta_0 T_{566}$.
Since $T_{566}$ is negative, a positive energy offset is required to obtain the appropriate sign of $R_{56}^{\mathrm{eff}}$. For example, with $T_{566}=-7.6$ mm and an energy offset of +1.5\%, the effective linear momentum compaction is $R_{56}^{\mathrm{eff}}\simeq -230~\mu$m. Importantly, the collimator was designed to support efficient FEL operation for beam energies offset by up to $\pm 1.5\%$ from the nominal value \cite{balandin}, making the off-energy compression scheme employed here feasible.

Third, the arc between SASE1 and SASE3 is used as a fifth compression stage (labeled 5 in Fig.~\ref{XFEL_layout}). The beam propagating through SASE1 accumulates an additional energy chirp due to the resistive-wall wakefield \cite{bane}, which is significant in the long ($\simeq 200$ m) and narrow vacuum chamber of the undulator. The mean beam energy is simultaneously reduced due to energy loss, bringing it closer to the design value after SASE1. To achieve strong compression in the arc, we detune a pair of quadrupoles from their nominal settings such that $R_{56}$ is negative, with an absolute value of a few hundred $\mu$m. This also introduces transverse dispersion, which is beneficial for short-pulse generation \cite{guetg,funke}.

\begin{figure}[tb]

\includegraphics[width=0.48\textwidth]{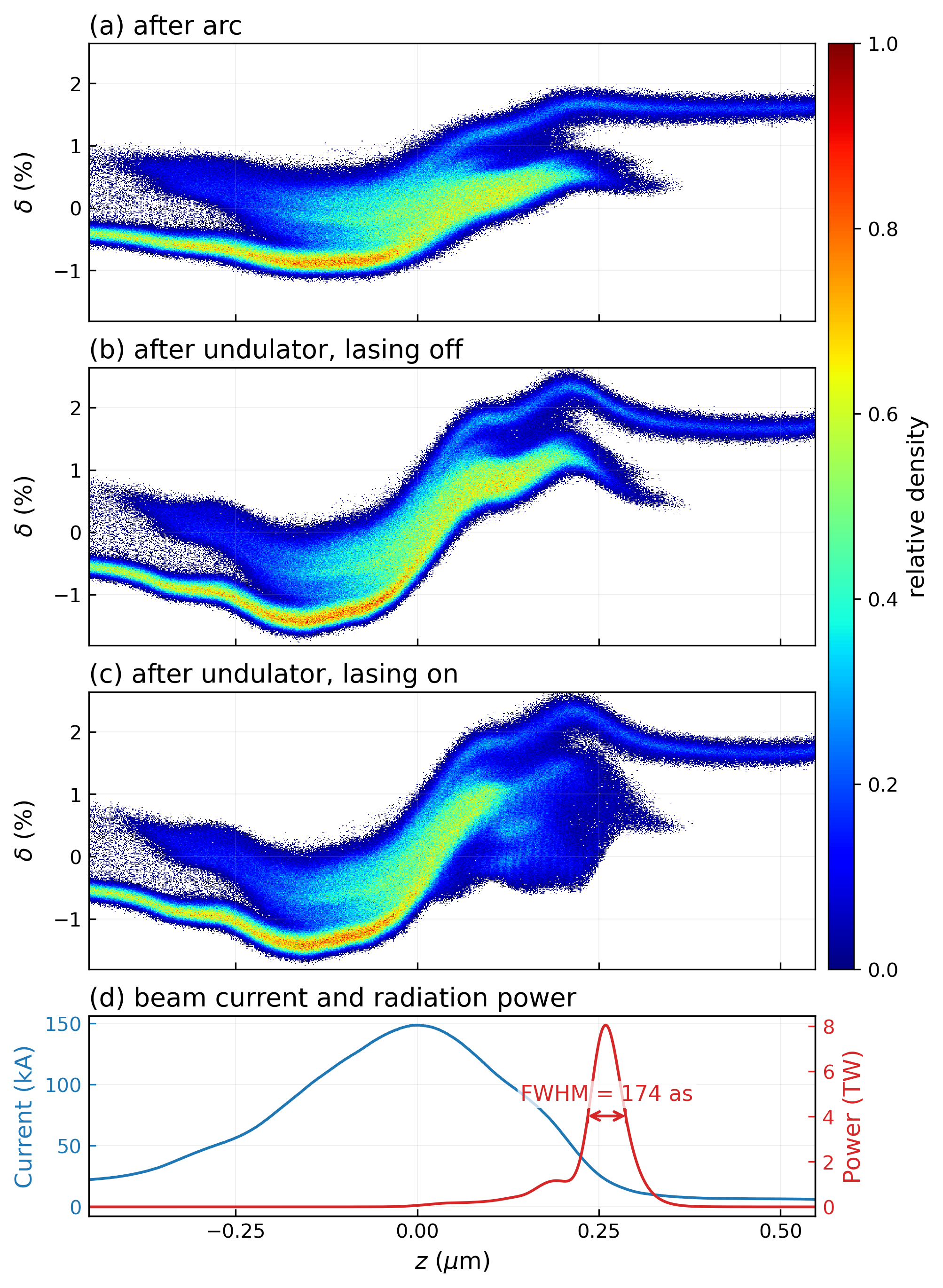}

\caption{
Simulated longitudinal phase space of the electron bunch (a) after the arc, (b) after the undulator with lasing suppressed, and (c) after the undulator with lasing enabled. The strong longitudinal space-charge effect in the undulator further modifies the phase space, while FEL interaction produces a localized distortion in the lasing region. (d) Beam-current profile (blue) and FEL power profile (red) for the shot shown in (c). The bunch head is on the right.}
\label{Phase_space}
\end{figure}

The beam dynamics simulations were performed using the codes Astra \cite{astra} and Ocelot \cite{ocelot}. The simulated longitudinal phase space after the arc is shown in Fig.~\ref{Phase_space}a. 
The beam is compressed to sub-$\mu$m scale, and the peak current reaches 150 kA (see Fig.~\ref{Phase_space}d), which is exceptionally high for x-ray FELs. 
Although this value cannot be directly confirmed experimentally in the SASE3 line of the European XFEL, a peak current of approximately 100 kA has recently been reported at SLAC National Accelerator Laboratory \cite{emma}.

A short electron bunch with an extremely high peak current remains subject to strong longitudinal space-charge (LSC) effects even at a beam energy of 14 GeV. These effects are particularly pronounced in undulators with large $K$ values \cite{rad-inter,LSC-und}, where the LSC impedance exceeds that of a drift space by a factor of $(1+K^2/2)$. The simulated longitudinal phase space downstream of the undulator in the absence of lasing is shown in Fig.~\ref{Phase_space}b. The LSC interaction in the undulator was modeled using the FEL code Genesis \cite{genesis}, which was also employed for simulations of the FEL amplification process. When lasing at a photon energy of 1 keV is included, the interaction with the radiation field produces a pronounced distortion of the longitudinal phase space at the bunch head, as shown in Fig.~\ref{Phase_space}c.

Our simulations demonstrate the advantage of generating soft x-rays with the high-energy electron beam available at the European XFEL. At other x-ray FEL user facilities, soft-x-ray undulators are typically driven by electron beams with energies of only 3--4 GeV. Even if electron bunches with properties similar to those shown in Fig.~\ref{Phase_space} could be generated, the resulting relative energy spread would be sufficiently large to prevent FEL amplification, while beam transport itself would become extremely challenging.

Apart from the longitudinal dynamics, an important aspect of bunch compression is the impact of collective effects in dispersive sections on the transverse beam dynamics. In particular, strong CSR for short, high-current bunches can lead to transverse emittance dilution \footnote{The effect on normalized emittance scales as $\gamma^{-1}$ or $\gamma^{-1/2}$ depending on the regime}. In our simulations, we indeed observe an increase of the core slice emittance from $\epsilon_n \simeq 0.6~\mu$m in the injector to $\epsilon_n \simeq 1~\mu$m in the undulator.
However, due to the large relativistic factor $\gamma$, the corresponding geometric emittance remains very small. As a result, the condition $4\pi \epsilon / \lambda < 1$ is well satisfied, owing to the operation of the soft-x-ray undulator at high electron energy.

We note that the use of resistive-wall wakefield for compression in a transfer line  was originally proposed in \cite{schlarb}. Later, the importance of LSC for the dynamics of a nonlinearly compressed beam in the linac of the first short-wavelength FEL facility FLASH, as well as the use of the LSC-induced energy chirp for compression in the dogleg (with negative $R_{56}$), was discussed in \cite{ttf-lsc}. This compression mode was routinely used during the first years of FLASH operation \cite{fl2006,fl2007}, leading to the generation of short extreme ultraviolet pulses. Subsequently, a similar concept was adapted \cite{lcls-nonlinear} for the generation of attosecond pulses in the x-ray regime and demonstrated experimentally \cite{lcls-hard,yan}. 
The use of second-order longitudinal dispersion for compression in combination with off-energy operation was explored in the VISA FEL experiments [40].
In the present work, we combine all of these techniques to achieve extremely strong compression and generate exceptionally powerful and short x-ray pulses.




Three experimental runs were performed in 2025 (run \#1) and 2026 (run \#2 and run \#3), yielding similar attosecond-pulse performance. Electron bunches with charges of 350--400 pC were generated and subjected to linearized full compression in the last standard bunch compressor, BC2 (labeled 3 in Fig.~\ref{XFEL_layout}). 
The beam was accelerated to 14.3 GeV, 300 MeV above the nominal energy for which the downstream beam optics was set, and subsequently compressed in the dogleg-collimator.
The high-frequency component of the signal from the bunch-length spectrometer \cite{nir} was used as an optimization observable and maximized. Finally, a pair of quadrupoles in the arc was detuned from its nominal setting to further compress the beam and optimize FEL performance.

The generation of attosecond pulses at a photon energy of 1.0 keV ($\lambda \simeq 1.2$ nm) took place in the SASE3 undulator ($\lambda_{\mathrm{w}}$ = 6.8 cm, $K=7.2$). To mitigate the effect of the strong energy chirp on FEL performance, we employed the chirp-taper compensation technique \cite{chirp-taper}. In this scheme, the undulator parameter $K$ is varied along the undulator to compensate for the energy variation experienced by the radiation pulse as it propagates through the chirped electron beam. The compensation also accounts for the small but finite $R_{56}$ of the undulator, which causes compression or decompression of the electron beam and, consequently, a change in the microbunching period.
Using the compensation condition derived in Ref.~\cite{chirp-taper}, one finds that the exceptionally strong taper employed in our experiment, corresponding to an increase  $\simeq 0.1\%/\mathrm{m}$, implies an energy chirp of order $1~\mathrm{GeV}/\mu\mathrm{m}$, consistent with the simulated phase space shown in Fig.~\ref{Phase_space}.

\begin{figure}[tb]

\includegraphics[width=0.45\textwidth]{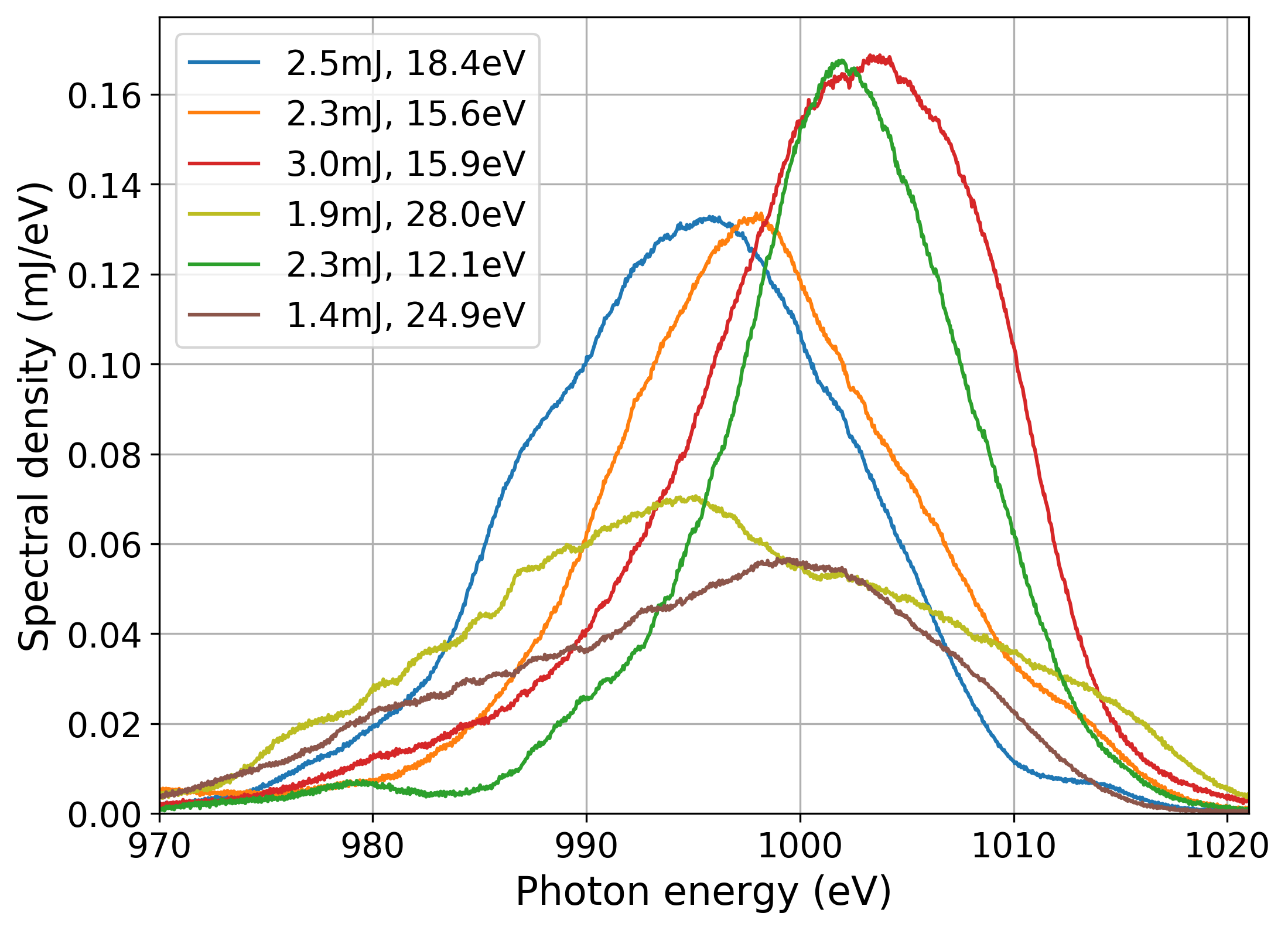}

\caption{
Examples of single-mode spectra of intense x-ray pulses. The legend shows pulse energy and FWHM bandwidth of each pulse.
}
\label{Spectra1}
\end{figure}

For the experiments reported here, we employed two photon diagnostic devices \cite{gruenert}. The spectra were measured with a grating spectrometer \cite{spectr} operating in the SASE3 beamline, providing an energy resolution of 0.35 eV. The pulse energy of individual x-ray pulses was measured with the non-invasive x-ray gas monitor (XGM) \cite{sorokin,theo}. Its operating principle is based on the photoionization of rare-gas atoms, with the resulting ion current absolutely calibrated to the photon flux. 
A strong linear correlation was observed between the pulse energy measured by the XGM and the spectrally integrated intensity recorded by the spectrometer, indicating that the XGM operates within the linear-response regime.



\begin{figure}[tb]

\includegraphics[width=0.45\textwidth]{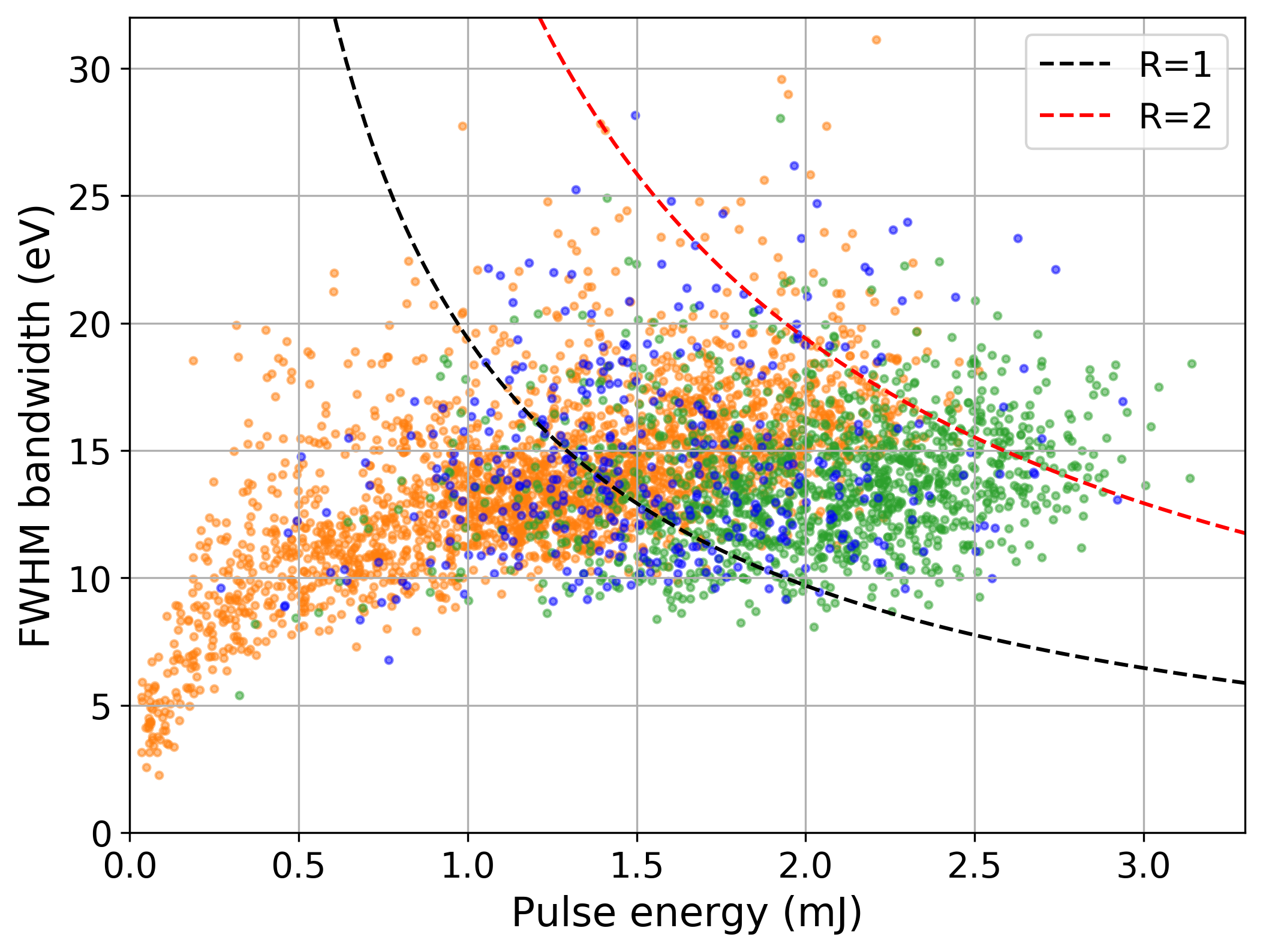}

\caption{Bandwidth versus pulse energy for single-mode events in run \#1 (blue dots), run \#2 (orange dots), and run \#3 (green dots). Dashed lines correspond to a peak power of 10 TW for $R=1$ (black) and $R=2$ (red), where $R$ is the ratio of the time-bandwidth product to its transform-limited value [Eq.~(\ref{rtl})].
}
\label{energy_BW}
\end{figure}

For the three experimental runs, the fraction of single-mode events was 38\%, 39\%, and 20\%, with corresponding average pulse energies of 1.6, 1.2, and 2.0 mJ, respectively. 
Examples of single-mode spectra are shown in Fig.~\ref{Spectra1}, illustrating individual pulses with FWHM bandwidths ranging from 10 to 30 eV and pulse energies up to 3 mJ.
The correlation between the pulse energy and the spectral width of the single-mode pulses is presented in Fig.~\ref{energy_BW}. These measurements demonstrate attosecond x-ray pulses with unprecedented millijoule-level pulse energies.

Metrology of attosecond x-ray pulses is developing rapidly and currently includes both temporal methods, such as angular streaking \cite{ang-streaking,duris-lcls-soft,franz-lcls-soft,funke}, and spectral methods. In this work, we rely on spectral measurements of single-mode SASE pulses, following approaches similar to those used in previous studies \cite{lcls-hard,duris-lcls-soft,swiss,yan}.

Since spectral measurements provide access only to the spectral intensity and not the spectral phase, additional assumptions are required to estimate the pulse duration. For example, the model proposed in \cite{lcls-hard} assumes a linear time-frequency correlation (chirp) and predicts that, for single-mode events in a high-gain FEL, the time-bandwidth product cannot exceed the Fourier (transform) limit by more than a factor of $\sqrt{2}$. This approach was subsequently used to estimate pulse durations in Refs.~\cite{lcls-hard,swiss,yan}.
Simultaneous measurements of attosecond x-ray pulses in the spectral and temporal domains using angular streaking \cite{lcls-streaking,funke}, as well as the recently developed double-blind holography technique \cite{agata}, indicate that the time-bandwidth product of individual single-mode SASE pulses typically exceeds the transform limit by a factor between one and two, with values close to unity being the most common.

For a pulse with a Gaussian intensity profile, the pulse duration can be expressed as

\begin{equation}
\Delta t (\mathrm{as}) =
\frac{1823 \times R}{\Delta E (\mathrm{eV})},
\label{rtl}
\end{equation}

\noindent where $\Delta t$ and $\Delta E$ are the FWHM pulse duration and energy bandwidth, respectively, and $R$ is the ratio of the time-bandwidth product to its transform-limited value. For example, a transform-limited pulse ($R=1$) with an energy bandwidth of 18 eV corresponds to a pulse duration of approximately 100 as. To guide the interpretation of Fig.~\ref{energy_BW}, we include two dashed lines corresponding to a peak power of 10 TW for $R=1$ and $R=2$. Of all events shown in Fig.~\ref{energy_BW}, 57\% lie above the $R=1$ line and 6\% above the $R=2$ line. These data provide strong evidence for the generation of attosecond pulses with peak powers on the order of 10 TW.

Despite the strong shot-to-shot fluctuations inherent to the SASE mechanism, future user experiments can preferentially select the shortest and most intense x-ray pulses using non-destructive pulse-energy measurements together with spectral diagnostics. Owing to the high repetition rate of the European XFEL, providing up to tens of thousands of pulses per second \cite{winni}, statistically significant data sets can be accumulated within practical measurement times.

In conclusion, we have demonstrated a source of soft-x-ray pulses with a duration on the order of 100 as and peak power in the 10-TW regime at the European XFEL. These results substantially extend the accessible parameter space of attosecond x-ray sources by combining ultrashort pulse duration with unprecedented peak power. Such pulses open new opportunities for ultrafast x-ray science, including studies of nonlinear phenomena and strongly driven electronic dynamics.

\section{Acknowledgments}

This paper is dedicated to the memory of Vladimir Balandin, whose invaluable contributions to the design of the European XFEL — particularly the innovative dogleg-collimator optics design — were foundational to the present work.
We acknowledge the support from the European XFEL GmbH, Schenefeld, and DESY, Hamburg, Germany.
This work was partly funded by the European XFEL research and development programme, beam time was provided
within the European XFEL facility development programme, and computational resources were obtained from DESY’s Maxwell cluster.
M.I. and T.M. acknowledge funding of the BMFTR-ErUM-Pro project “AttoSee” (05K25GU4). M.I. and K.D. acknowledge funding of the BMFTR-ErUM-Data project “OPTIX” (05D2025). M.I. acknowledges funding from the Deutsche Forschungsgemeinschaft - project "CACTUS" (707468). L.F., S.S., K.S., L.W., M.I. and W.H. acknowledge funding of the BMBF-ErUM-Pro project “TRANSALP” (05K22PE3) and the BMBF project “SpeAR XFEL” (05K19PE1).

\clearpage

\bibliographystyle{apsrev4-2}
\bibliography{atto_10tw_prl_full}

\end{document}